\documentstyle[amstex,preprint,prl,eqsecnum,aps]{revtex}

\begin{document}

\title{Wigner Distribution Function Approach to
Dissipative Problems in Quantum Mechanics with emphasis on
Decoherence and Measurement Theory}

\author{{R.F. O'Connell}  \\ {\it{Department of Physics
and Astronomy, Louisiana State University, \\ Baton
Rouge, LA  70803-4001}}}

\date{\today}

\maketitle

\begin{abstract} We first review the usefulness of the
Wigner distribution functions (WDF), associated with
Lindblad and pre-master equations, for analyzing a host of
problems in Quantum Optics where dissipation plays a major
role, an arena where weak coupling and long-time
approximations are valid.  However, we also show their
limitations for the discussion of decoherence, which is
generally a short-time phenomenon with decay rates
typically much smaller than typical dissipative decay
rates.  We discuss two approaches to the problem both of
which use a quantum Langevin equation (QLE) as a
starting-point:  (a) use of a reduced WDF but in the
context of an exact master equation (b) use of a WDF for
the complete system corresponding to entanglement at all
times. 
\end{abstract}


\newpage

\section{Introduction}

In recent years, there has been widespread interest in
dissipative problems arising in a variety of areas in
physics.  Historically, particularly in quantum optics and
NMR (where weak coupling and long-time approximations are
valid), master equations for the time evoluation of the
density matrix have been extensively utilized in
analyzing such
problems\cite{louisell73,lindblad76,caldeira83,dekker77,walls94,ford96}
and it has also being found useful to re-express such
equations in terms of the Wigner distribution
function\cite{louisell73,dekker77}.  However, with the
recent burgeoning interest in decoherence, which is
generally, a short-time phenomenon with decay rates much
smaller than typical dissipative decay rates, a new
element arises.  Recognizing that conventional master
equation approaches are often not adequate, some
investigators have used path integral
methods\cite{caldeira83}.  However, we have found that
the simplest and most physically appealing approach to the
problem is via use of generalized quantum Langevin
equations\cite{ford88}, supplemented by use of Wigner
distribution functions
\cite{hillery84}.  Our focus is on quantum Brownian
motion: a quantum particle interacting with a heat bath
and moving in an arbitrary external potential.  For the
investigation of decoherence phenomena, we have found
\cite{ford01,ford012} that it is important to distinguish
between two different physical scenarios viz. (a) complete
entanglement between the quantum particle and the heat
bath at all times and (b) the system in a state in which
the oscillator is not coupled to the bath at, say, $t=0$
and such that the bath is in equilibrium at temperature
$T$.  Thus, it takes a characteristic time of the order of
$\gamma^{-1}$ (where $\gamma$ is a typical dissipative
decay time) for the complete coupling to occur and for the
whole system to come into thermal equilibrium.

Scenario (b), with the assumption of an uncoupled initial
state is that used in the derivation of master equations,
including the exact master equation of Hu, Paz and Zhang
\cite{ford012}.  However, this scenario is inappropriate
for the calculation of very short decoherence decay times
and, in general, one must abandon master equation methods
for this purpose.  For analysis of scenario (a), by its
nature, it is necessary to abandon master equation
methods.  Instead, we have employed
\cite{ford01} a more general method due to Ford and Lewis
\cite{ford86}.  Finally, the presence of a classical
external force $f(t)$ brings a new dimension to the
problem.

The familiar master equation of quantum optics is in
Lindblad form\cite{lindblad76}, which guarantees that the
density matrix is always positive definite during time
evolution.  In the derivation of this equation
\cite{louisell73,ford96}, rapidly oscillating terms are
omitted by the method of coarse-graining in time; the high
frequencies correspond to the oscillator frequency
$\omega_0$ and, in the usual weak coupling limit,
${\omega_0}>>{\gamma}$, where $\gamma$ is a typical decay
constant.

We have referred to the equations obtained prior to
coarse-graining in time as pre-master (or pre-Lindblad)
equations \cite{ford96,ford99} and such equations have
been used extensively in other areas of physics
\cite{caldeira83,dekker77}; other authors have simply
referred to them as master equations but, to avoid
confusion, we reserve the latter term for equations in
Lindblad form.  Pre-master equations, like the master
equation, describe an approach to the equilibrium state.
This equilibrium state is the same in either
case\cite{ford99}, but with pre-master (non-Lindblad)
equations the approach can be through non-physical states
of negative probability.  However, as we demonstrate here,
pre-master equations have other advantages
vis-${\grave{a}}$-vis master equations: (a) they lead to
the exact expression for the mean value of $x(t)$ (as
obtained from the exact Langevin equation for the
problem), (b) they lead, in the classical limit
$(\hbar\rightarrow 0)$, to the familiar Fokker-Planck
equation of classical probability, (c) the exact master
equation
\cite{hpz92} is for long times of pre-master form,
positivity being preserved through the time dependence of
the coefficients.  However, the pre-master equations can
exhibit different forms, depending on whether momentum or
coordinate coupling forms of
$H$ have been used.

It is thus clear that there are many facets to the overall
question of analyzing dissipative systems.  Our purpose is
to delineate the strengths and weaknesses of the various
approaches.  As will become apparent, one common thread
which will permeate the discussion is the usefulness of
the Wigner distribution for all scenarios.

As already indicated, there is an important dividing line
between long-time $(t>>\gamma^{-1})$ and short-time
$(t<<\gamma^{-1})$ phenomena.  Thus, in Sec. II, we
discuss the quantum Langevin equation for scenario (a),
which describes quantum particle and bath entanglement at
all times, and we consider in turn both the stationery
process (corresponding to the absence of an external force
$f(t)$ in which case we are dealing with the system in
thermal equilibrium) and then the case of $f(t)\neq 0$.
This naturally leads us to a comparison of the Langevin
equations of motion with those derived (in the Ohmic limit
and for an oscillator) using the conventional master
equation techniques of Quantum Optics and NMR.  Thus, as a
starting-point, we write down in Sec. III, the quantum
Langevin equation for the Ohmic model.  In addition, we
present, in the interaction representation, the two forms
(momentum coupling and coordinate coupling) of the
weak-coupling pre-master equations as well as the
corresponding master equation (which is the same for both
couplings\cite{ford99}).

Then in Sec. IV, we consider in detail the momentum and
coordinate coupling forms, respectively, of the pre-master
equations as well as the master equation.  In Sec. V, we
discuss the corresponding Wigner quantum probability
distributions and their classical limit.

To incorporate short-time phenomena, a more sophisticated
analysis is required, requiring both the Langevin equation
discussed in Sec. II as well as a generalization to a
Langevin equation for the initial value problem, which is
the theme of Sec. VI.  First, we present the solution of
this equation in terms of the solution of the stationery
equation, the corresponding Green function and the
fluctuation force.  Next, we write this solution in a form
local in time.  Moreover, we point out that the equivalent
master equation, as distinct from those discussed in
Secs. III to V, is exact in that the coefficients are
time-dependent for which explicit expressions are readily
determined using the aforementioned solution of the
Langevin equation.  Then, in Sec. VII, with a view toward
calculation of quantum probability functions for
superposition states and the corresponding decay of the
interference term (decoherence) and developing a
framework which will accomodate many different
possibilities, we are led again to a consideration of
Wigner distribution functions, both for the whole system
of quantum particle and bath and also (after averaging
over the bath variables) for the reduced system.  We also
show that it is sometimes useful to express the solution
for the Wigner function at time $t$ in the form of an
integral of a transition probability acting on the initial
Wigner function; in particular, we are led to a
generalization of the classical Kramers equation, in the
sense that it is quantum, applies to an arbitrary heat
bath and allows for a time-dependence in the external
force.  In Sections VI and VII, we confined ourselves to
$f(t)=0$ and, next, in Sec. VIII, we show how these
various results can be applied to the calculation of
decoherence, both for the case of (a) entanglement at all
times and (b) the initial value problem where the system
and bath are initially decoupled.  In addition to our
detailed discussion of decoherence in coordinate space, we
also comment on results obtained for decoherence in both
phase space and momentum space.  The case of $f(t)\neq 0$
is also considered.

Figs. 1 and 2 present an overview of the paper which
should hopefully enable the reader to "see the wood from
the trees."  In particular, in Fig. 1, we see that
specification of the complete Hamiltonian $H$ leads
immediately to a quantum Langevin equation, the exact form
for which depends on whether one is considering
entanglement at all times or an initial value problem.
From this equation, one is led to Wigner quantum
probability distributions from which one may obtain, in
particular, probabilities in either coordinate space or
momentum space.  These are the tools necessary for
consideration of various problems such as decoherence
phenomena.  Fig. 2 deals with the weak coupling limit of
exact master equations and delineates their shortcomings.

\section{Generalized Quantum Langevin Equation:
Entanglement for all Time}

In recent years, there has been widespread interest in
dissipative problems arising in a variety of areas in
physics.  As it turns out, solutions of many of these
problems are encompassed by a generalization of Langevin's
equation to encompass quantum, memory, and non-Markovian
effects, as well as arbitrary temperature and the presence
of an external potential $V(x)$.  As in Ref.
\cite{ford88}, we refer to this as the generalized quantum
Langevin equation (QLE)

\begin{equation} m \ddot{x}
+\int^{t}_{-\infty}dt^{\prime}\mu
(t-t^{\prime})\dot{x}(t^{\prime})+V^{\prime}(x)=F(t)+f(t),
\label{qle1}
\end{equation} where $V^{\prime}(x)=dV(x)/dx$ is the
negative of the time-independent external force and $\mu
(t)$ is the so-called memory function.  $F(t)$ is the
random(fluctuation or noise) force and
$f(t)$ is a $c$-number external force.  In addition, it
should be strongly emphasized that "-- the description is
more general than the language --" \cite{ford88} in that
$x(t)$ can be a generalized displacement operator (such as
the phase difference of the superconducting wave  function
across a Josephson junction).

A detailed discussion of (\ref{qle1}) appears in Ref.
\cite{ford88}.  In particular, it was pointed out the QLE
corresponds to a macroscopic description of a quantum
system interacting with a quantum-mechanical heat-bath
and that this description can be precisely formulated,
using such general principles as causality and the second
law of thermodynamics.  We also stressed that this is a
model-independent description.  However, the most general
QLE can be realized with a simple and convenient model,
viz., the independent-oscillator (IO) model.  The
Hamiltonian of the IO system is

\begin{equation}
H=\frac{p^{2}}{2m}+V(x)+\sum_{j}\left(\frac{p^{2}_{j}}{2m_{j}}+\frac{1}{2}m_{j}\omega^{2}_{j}(q_{j}-x)^{2}
\right)-xf(t). \label{qle2}
\end{equation} Here $m$ is the mass of the quantum
particle while $m_{j}$ and $\omega_{j}$ refer to the mass
and frequency of heat-bath oscillator $j$.  In addition,
$x$ and $p$ are the coordinate and momentum operators for
the quantum particle and $q_{j}$ and $p_{j}$ are the
corresponding quantities for the heat-bath oscillators.

The infinity of choices for the $m_{j}$ and $\omega_{j}$
give this model its great generality\cite{ford88}.  In
particular, it can describe nonrelativistic quantum
electrodynamics\cite{ford88}, the Schwabl-Thirring (S-T)
model\cite{schwabl64}, the Ford-Kac-Mazur (FKM)
model\cite{ford65}, and the Lamb model
\cite{lamb00}.

In this context, it should be noted that, whereas $H$ in
(\ref{qle2}) has been put into a form in which all the
heat-bath oscillators interact with the central
oscillator of interest but not with each other, we have
shown that this is the most general $H$ one can write down
to describe most types of dissipation encountered in the
literature; in particular, it is unitarily equivalent to
the FKM model in which all the oscillators are coupled (as
discussed in Sec. VE of Ref. \cite{ford88}).  Also, of
interest, in connection with our later discussions, is the
velocity-coupling model in which the coupling is through
the particle momentum (as distinct from the particle
coordinate as in (\ref{qle2})), corresponding to the
Hamiltonian

\begin{equation}
H_{VC}=\frac{1}{2m}\left(p+\sum_{j}m_{j}\omega_{j}q_{j}\right)^{2}+V(x)+\sum_{j}\left(\frac{p^{2}_{j}}
{2m_{j}}+\frac{1}{2}m_{j}\omega^{2}_{j}q^{2}_{j}\right)-xf(t).
\label{qle3}
\end{equation} As shown explicitly in \cite{ford88},
$H_{VC}$ is obtained from $H$ by means of two unitary
transformations which change the
$p,~p_{j},~{\textnormal{and}}~q_{j}$ but which do not
change the coordinate $x$.  We have refrained from
putting primes on the changed variables since the context
will always ensure that no confusion will arise.
However, for future purposes, we note that, in proceeding
from (\ref{qle2}) to (\ref{qle3}),

\begin{equation} p\rightarrow
p-\sum_{j}m_{j}\omega_{j}q_{j}. \label{qle4}
\end{equation} Thus, it is not surprising that both the IO
model and the vector coupling model lead to the same
identical GLE (\ref{qle1}).  Morover, since both
Hamiltonians (\ref{qle2}) and (\ref{qle3}) are unitarily
equivalent to the Hamiltonian of non-relativistic quantum
electrodynamics (QED) we see the close analogy with the
equivalence of the $\vec{x}\cdot\vec{E}$ and
$\vec{p}\cdot\vec{A}$ couplings in QED.  Whereas most of
our previous work has used the IO form of the Hamiltonian,
we found it convenient in
\cite{ford96} to use the VC form for the derivation of a
non-Markovian master equation.

Use of the Heisenberg equations of motion leads to the
GLE(2.1) describing the time development of the particle
motion, where

\begin{equation}
\mu(t)=\sum_{j}m_{j}\omega^{2}_{j}\cos(\omega_{j}t)\theta(t),
\label{qle5}
\end{equation} is the memory function, with $\theta (t)$
the Heaviside step function.  Also

\begin{equation}
F(t)=\sum_{j}m_{j}\omega^{2}_{j}q^{h}_{j}(t), \label{qle6}
\end{equation} is a fluctuating operator force with mean
$\langle F(t)\rangle =0$, where $q^{h}(t)$ denotes the
general solution of the homogeneous equation for the
heat-bath oscillators (corresponding to no interaction).
These results were used to obtain the results for the
(symmetric) autocorrelation and commutator of $F(t)$,
viz.,

\begin{eqnarray}
\frac{1}{2}\langle
F&&(t)F(t^{\prime})+F(t^{\prime})F(t)\rangle \nonumber \\
&&~= \frac{1}{\pi}\int^{\infty}_{0}d\omega
Re[\tilde{\mu}(\omega+i0^{+})]\hbar\omega\coth
(\hbar\omega/2kT)\cos[\omega(t-t^{\prime})] \label{qle7}
\end{eqnarray}

\begin{eqnarray} [F(t),F(t^{\prime})] &=&
\frac{2\hbar}{i\pi}\int^{\infty}_{0}d\omega
Re\{\tilde{\mu}(\omega +i0^{+})\}\omega
\nonumber \\ &&{}\times\sin\omega(t-t^{\prime}).
\label{qle33}
\end{eqnarray} Here $\tilde{\mu}(z)$ is the Fourier
transform of the memory function:

\begin{equation}
\tilde{\mu}(z)=\int^{\infty}_{0}dt\mu (t)e^{izt}.
\label{qle34}
\end{equation}

Equation (\ref{qle7}) is referred to by Kubo \cite{kubo66}
as the second fluctuation-dissipation theorem and we note
that it can be written down explicitly once the GLE is
obtained.  Also, its evaluation requires only knowledge
of Re$\tilde{\mu}(\omega)$.  On the other hand, the first
fluctuation-dissipation theorem is an equation involving
the autocorrelation of $x(t)$ and its explicit evaluation
requires a knowledge of the generalized susceptibility
$\alpha(\omega)$ (to be defined below) which is equivalent
to knowing the solution to the GLE and also requires
knowledge of both Re$\tilde{\mu}(\omega)$ and
Im$\tilde{\mu}(\omega)$.  This solution is readily
obtained when $V(x)=0$, corresponding to the original
Brownian motion problem \cite{langevin08}.  As shown by
FLO \cite{ford88}, a solution is also possible in the case
of an oscillator.  Taking
$V(x)=\frac{1}{2}Kx^{2}=\frac{1}{2}m\omega^{2}_{0}x^{2}$,
these authors obtained the solution of the  Langevin
equation (\ref{qle1}) in the form
\begin{equation} x(t)=\int_{-\infty }^{t}dt^{\prime
}G(t-t^{\prime })\{F(t^{\prime })+f(t^{\prime})\},
\label{qle35}
\end{equation} where $G(t)$, the Green function, is given
by

\begin{equation} G(t)=\frac{1}{2\pi }\int_{-\infty
}^{\infty }d\omega \alpha (\omega +i0^{+})e^{-i\omega
t},  \label{qle36}
\end{equation} with $\alpha (z)$ the familiar response
function

\begin{equation}
\alpha (z)=\frac{1}{-mz^{2}-iz\tilde{\mu}(z)+K}.
\label{qle37}
\end{equation} It is often convenient to write
(\ref{qle35}) in the form

\begin{equation} x(t)=x_{s}(t)+x_{d}, \label{qle64}
\end{equation} where $x_{d}$ is the "driven" contribution
\cite{ford00} due to the external force $f(t)$ and $x_{s}$
is the contribution due to the fluctuation force $F(t)$.
Here we have introduced a subscript s to emphasize that
$x_{{\rm s}}(t)$ is a stationary operator-process, in the
sense that correlations, probability distributions, etc.
for this dynamical variable are invariant under
time-translation ($t\rightarrow t+t_{0}$). In particular,
the correlation,

\begin{eqnarray}
C_{o}(t-t^{\prime})\equiv\frac{1}{2}\langle x_{{\rm
s}}&&(t)x_{{\rm s}}(t^{\prime })+x_{{\rm s}
})x_{{\rm s}}(t)\rangle \nonumber \\ &&{}=\frac{\hbar
}{\pi }%
\int_{0}^{\infty }d\omega {\rm Im}\{\alpha (\omega
+i0^{+})\}\coth \frac{%
\hbar \omega }{2kT}\cos \omega (t-t^{\prime }),
\label{qle38}
\end{eqnarray} is a function only of the time-difference
$t-t^{\prime }$.  Furthermore, \cite{ford00}

\begin{eqnarray} C_{d}(t,t^{\prime}) &\equiv&
\frac{1}{2}\langle
x(t)x(t^{\prime})+x(t^{\prime})x(t)\rangle \nonumber \\
&=& C_{o}(t-t^{\prime})+\langle x(t)\rangle \langle
x(t^{\prime})\rangle, \label{qle65}
\end{eqnarray} where $\langle x(t)\rangle$ is the steady
mean of the driven motion.

Also, taking the Fourier transform of (\ref{qle35}), we
obtain

\begin{equation}
\tilde{x}(\omega)=\alpha(\omega)\{\tilde{F}(\omega)+\tilde{f}(\omega)\},
\label{qle8}
\end{equation} where the superposed tilde is used to
denote the Fourier transform.  Thus, $\tilde{x}(\omega)$
is the Fourier transform of the operator $x(t)$:

\begin{equation}
\tilde{x}(\omega)=\int^{\infty}_{-\infty}dtx(t)e^{i\omega
t}. \label{qle10}
\end{equation}

We have now all the tools we need to calculate observable
quantities.

\section{Quantum Langevin equation for the Ohmic model and
Corresponding Weak Coupling Master Equations}

As a special case of (\ref{qle1}), we see that in the
particular case of an oscillator with spring constant
$K$, the Langevin equation for Ohmic coupling (constant
$\gamma$), has the form

\begin{equation} m{\ddot{x}}+m
\gamma{\dot{x}}+Kx=F(t)+f(t), \label{qle12}
\end{equation} Moreover,

\begin{equation}
\langle{\dot{x}}(t)\rangle={\frac{{\langle}p(t)\rangle}{m}},\qquad
\langle\dot{p}(t)\rangle=-{\gamma}\langle{p}(t)\rangle-K\langle{x}(t)\rangle
+f(t). \label{qle13}
\end{equation}

These are exact results.  In the weak coupling limit, the
corresponding pre-master and master equations for the
density matrix may be obtained.  First, a word about
notation\cite{ford96}.  In the Schr\"{o}dinger
representation, the density matrix for the whole system
(quantum oscillator plus heat bath) is denoted by $\rho$
whereas $\tilde{\rho}$ denotes the reduced density matrix,
obtained after a partial trace has been carried out over
the heat bath variables (which, in effect, eliminates the
very rapid stochastic fluctuations).  The equation for
$\tilde{\rho}$ is what we refer to as a pre-master
equation and it includes high frequency oscillating terms
$\exp\{\pm2i\omega_{0}t\}$, where $\omega_{0}$ is the
oscillator frequency.  If the equation for
$\tilde{\rho}$ is time-averaged (which is equivalent to
discarding the high-frequency oscillating terms) then the
resulting equation is called a master equation for the
time-averaged density matrix which is denoted by
$\bar{\rho}$.  The corresponding quantities in the
interaction representation are
$\sigma$, $\tilde{\sigma}$ and $\bar{\sigma}$.
Explicitly \cite{ford96}, in the interaction
representation,

\begin{equation}
\sigma=\exp\left\{\frac{iHt}{\hbar}\right\}\rho\exp\left\{-\frac{iHt}{\hbar}\right\},
\label{qle55}
\end{equation}

\begin{equation}
\tilde{\sigma}(t)=Tr_{R}\{\sigma (t)\}, \label{qle56}
\end{equation} (where $Tr_{R}$ refers to the trace over
the reservoir),

\begin{equation}
\frac{\partial\tilde{\sigma}}{\partial
t}=\frac{\partial\bar{\sigma}}{\partial t}+\exp\{\pm
2i\omega_{0}t\}~~{\textnormal{terms}}, \label{qle57}
\end{equation} and we have corresponding equations in the
Schr\"{o}dinger representation.

Moreover, the form of any of these density matrix
equations depends on whether one is using the momentum or
the coordinate coupling forms.  In the interaction
representation
\cite{ford96,ford99} the momentum coupling form can be
written

\begin{equation}
{\frac{{\partial}{\tilde\sigma}}{\partial{t}}}=-{\frac{\gamma}{2\hbar}}\lbrace{\frac{1}{m\omega_0}}
\coth{\frac{\hbar\omega_0}{2kT}}{\lbrack}p(t),
{\lbrack}p(t),
{\tilde\sigma}{\rbrack}{\rbrack}-i{\lbrack}p(t),
x(t){\tilde\sigma}+{\tilde\sigma}x(t){\rbrack}{\rbrace},
\label{qle14}
\end{equation} while the coordinate coupling form can be
written

\begin{equation}
{\frac{{\partial}{\tilde\sigma}}{\partial{t}}}=-{\frac{\gamma}{2\hbar}}{\lbrace}m{\omega_0}\coth
{\frac{\hbar{\omega_0}}{2kT}}{\lbrack}x(t),
{\lbrack}x(t),{\tilde\sigma}{\rbrack}{\rbrack}+i{\lbrack}x(t),p(t){\tilde\sigma}+{\tilde\sigma}p(t){\rbrack}
{\rbrace}. \label{qle15}
\end{equation}

From either of these equations one obtains the same master
equation by discarding the terms explicitly oscillating at
frequency $2\omega_{0}$ (often referred to as the
rotating-wave approximation (RWA)) and this is best
achieved by transforming from the variables $x,p$ to $a$
and $a^{+}$\cite{ford96}.

The corresponding master equation is
\cite{ford99}

\begin{eqnarray}
{\frac{{\partial}{\bar{\sigma}}}{\partial{t}}}=-{\frac{\gamma}{4\hbar}}{\lbrace}\coth
{\frac{\hbar{\omega_0}}{2kT}}  &&
({\frac{1}{m{\omega_0}}}{\lbrack}p,{\lbrack}p,\bar{\sigma}{\rbrack}{\rbrack}+m{\omega_0}{\lbrack}x,{\lbrack}
x,\bar{\sigma}{\rbrack}{\rbrack}){\nonumber} \\ &&
+i({\lbrack}x,p\bar{\sigma}+\bar{\sigma}p{\rbrack}-{\lbrack}p,x\bar{\sigma}+\bar{\sigma}x{\rbrack})
{\rbrace}. \label{qle16}
\end{eqnarray} and it is a simple matter to verify that
this has the Lindblad form of the master equation familiar
in quantum optics\cite{louisell73,walls94,ford96}.  On
the other hand, outside this community, the pre-master
equations have been used extensively by Caldeira and
Leggett\cite{caldeira83} and others.  We wish to
emphasize that all three of the above equations lead to
the same equilibrium state:
$\rho_{cq}=\exp\{-H_{0}/kT\}$; i.e., detailed balance is
obeyed.  The difference is in the {\emph{approach}} to
equilibrium.  For the pre-Lindblad equations this can be
through (unphysical) states in which $\rho$ is not
positive definite.  Note that this form of the
equilibrium state holds only if $H_{0}$ is the free
oscillator Hamiltonian (given below in (\ref{qle67})).

As already alluded to, whereas the Lindblad form of the
master equation has the merit of ensuring a positive
definite matrix element at all times (in contrast to
pre-master equations) it has the disadvantage of being
subject to the RWA.  On the other hand, there are
undesirable features associated with the non-positivity
associated with the pre-master
equations\cite{ambegaokar93} but it should be remarked
that all master equations give an incorrect form for very
short times, as can be shown most simply by calculating
the mean-square displacement, defined in (\ref{qle53})
below, by means of both an exact calculation and a master
equation approach.  On the other hand, after a short
initial time of the order of the "bath relaxation time"
(zero for the Ohmic case, $\tau$ for the single relaxation
time case) the exact equation has the pre-master form.
This can be seen from \cite{ford012}, where in equation
(\ref{qle28}) we gave explicit expressions for the Ohmic
case or in equation (A5) for the single relaxation time
case.  Another point is that the pre-master equation gives
the exact equation of motion for the mean motion, the
master equation does not.

\section{Mean values for the various forms of the density
matrix}

\subsection{The Momentum Coupling Form of the Pre-Master
Equation}

In the Schr\"{o}dinger representation (\ref{qle14})
becomes \cite{ford96}

\begin{equation}
{\frac{{\partial}{\tilde\rho}}{\partial{t}}}={\frac{1}{i\hbar}}{{\lbrack}H_{d},{\tilde\rho}}\rbrack-{\frac{\gamma}{2\hbar}}
{\lbrace}{\frac{1}{m{\omega_0}}}\coth{\frac{\hbar{\omega_0}}{2kT}}{\lbrack}p,{\lbrack}p,{\tilde\rho}
{\rbrack}{\rbrack}
-i{\lbrack}p,x{\tilde\rho}+{\tilde\rho}x{\rbrack}{\rbrace},
\label{qle17}
\end{equation} where $H_{d}$, the Hamiltonian for the
driven oscillator, is given by

\begin{equation} H_{d}=H_{o}-xf(t). \label{qle66}
\end{equation} and where

\begin{equation}
H_{o}=\frac{p^{2}}{2m}+\frac{1}{2}Kx^{2}, \label{qle67}
\end{equation} is the free oscillator Hamiltonian
\cite{ford00}.

The corresponding equations for the mean of $x$ and $p$ are

\begin{equation}
{\frac{d}{dt}}{\langle}x{\rangle}={\frac{{\langle}p{\rangle}}{m}}-\gamma{\langle}x{\rangle},\qquad
{\frac{d}{dt}}{\langle}p{\rangle}=-m{\omega^2_0}{\langle}x{\rangle}+f(t).
\label{qle18}
\end{equation}

If we eliminate ${\langle}p{\rangle}$ between these two
equations, we get

\begin{equation}
{\frac{{d^2}{\langle}x{\rangle}}{dt^2}}+{\gamma}{\frac{d{\langle}x{\rangle}}{dt}}+{\omega^2_0}
{\langle}x{\rangle}=f(t). \label{qle19}
\end{equation}

This, of course, is just what we would get from the
quantum Langevin equation.

\subsection{The Coordinate Coupling Form of the
Pre-Master Equation}

In the Schr{\"{o}}dinger representation (\ref{qle15})
becomes

\begin{equation}
{\frac{{\partial}{\tilde\rho}}{\partial{t}}={\frac{1}{i\hbar}}{\lbrack}H_{d},{\tilde\rho}{\rbrack}-
{\frac{\gamma}{2\hbar}
{\lbrace}m{\omega_0}{\coth}}{\frac{\hbar{\omega_0}}{2kT}}{\lbrack}x,{\lbrack}x,{\tilde\rho}\rbrack\rbrack
+i\lbrack x,p{\tilde\rho}+{\tilde\rho}p
{\rbrack}{\rbrace}}. \label{qle20}
\end{equation}

Here, and also below, we neglect the small energy shifts
because they are of no consequence as far as the present
discussion is concerned.  The corresponding equations for
the mean of $x$ and $p$ are

\begin{equation}
{\frac{d}{dt}}{\langle}x{\rangle}={\frac{{\langle}p{\rangle}}{m}},\qquad
{\frac{d}{dt}}{\langle}p{\rangle}=-m{\omega^2_0}{\langle}x{\rangle}-{\gamma}{\langle}p{\rangle}+f(t).
\label{qle21}
\end{equation}

If again we eliminate ${\langle}p{\rangle}$ between these
two equations, we get the same equation (\ref{qle19}) as
in the momentum-coupling case.  In other words, both forms
of the pre-master equation lead to what we would get from
the exact quantum Langevin equation.

\subsection{The Master Equation}

In the Schr{\"{o}}dinger representation (\ref{qle16})
becomes

\begin{eqnarray}
{\frac{{\partial}{\bar\rho}}{{\partial}t}={\frac{1}{i\hbar}}{\lbrack}H_{d},{\bar\rho}\rbrack-{\frac{\gamma}{4\hbar}}}
&&
{\lbrace}\coth{\frac{{\hbar}{\omega_0}}{2kT}}({\frac{1}{m{\omega_0}}}{\lbrack}p,{\lbrack}p,{\bar\rho}{\rbrack}
{\rbrack}+m\omega_0 \lbrack x, \lbrack
x,{\bar\rho}\rbrack\rbrack){\nonumber} \\ &&
+i({\lbrack}x,p{\bar\rho}+{\bar\rho}p{\rbrack}-{\lbrack}p,x{\bar\rho}+{\bar\rho}x{\rbrack}){\rbrace}.
\label{qle22}
\end{eqnarray}

The corresponding equations for the mean of $x$ and $p$
are\cite{ford00}

\begin{equation}
{\frac{d{\langle}x{\rangle}}{dt}}={\frac{{\langle}p{\rangle}}{m}}-{\frac{\gamma}{2}}{\langle}x{\rangle},
{\frac{d{\langle}p{\rangle}}{dt}}=-m{\omega^2_0}{\langle}x{\rangle}-{\frac{\gamma}{2}}{\langle}p{\rangle}+f(t). 
\label{qle23}
\end{equation}

Eliminating ${\langle}p{\rangle}$ between these two
equations, we get

\begin{equation}
{\frac{{d^2}{\langle}x{\rangle}}{dt^2}}+{\gamma}{\frac{d{\langle}x{\rangle}}{dt}}+({\omega^2_0}+{\frac{{\gamma^2}}{4}}
){\langle}x{\rangle}=\frac{f(t)}{m}. \label{qle24}
\end{equation}

Because of the spurious $({\gamma^2}/4)$ term, this is
{\underline{not}} the equation one gets from the quantum
Langevin equation.  In fact, this equation is identical
with the equation obtained from use of an RWA
approximation to the exact Hamiltonian \cite{ford962}.

\section{Quantum Probability Distributions} These are
most simply obtained by using the Wigner function which is
defined by the relation \cite{hillery84}

\begin{equation}
W(q,p;t)\equiv{\frac{1}{2\pi\hbar}}\int^{\infty}_{-\infty}due^{iup/\hbar}\rho(q-{\frac{u}{2}},q+{\frac{u}{2}};t). 
\label{qle25}
\end{equation}

The translation from the operator equation for $\rho$ to
the quasi-classical equation for the Wigner function is
most easily carried out by use of the Bopp operators
\cite{hillery84}, as a result of which we get the
correspondence

\begin{eqnarray} &&
x\rho\leftrightarrow(q+{\frac{i\hbar}{2}}{\frac{\partial}{\partial
p})W,
\qquad
\rho{x}\leftrightarrow(q-{\frac{i\hbar}{2}}}{\frac{\partial}{\partial
p}})W,{\nonumber} \\ &&
p\rho\leftrightarrow(p-{\frac{i\hbar}{2}}{\frac{\partial}{\partial
q}})W, \qquad \rho p\leftrightarrow(p+
{\frac{i\hbar}{2}}{\frac{\partial}{\partial q}})W.
\label{qle26}
\end{eqnarray}

It follows that the corresponding equation for the Wigner
function is

\begin{eqnarray} {\frac{\partial W}{\partial t}}= &&
-{\frac{p}{m}}{\frac{\partial W}{\partial
q}}+m{\omega^2_0}q{\frac{\partial W} {\partial
p}}-f(t)\frac{\partial W}{\partial p}{\nonumber} \\ &&
+{\frac{\gamma}{2}}\lbrace(1+\lambda){\frac{\partial
qW}{\partial q}}+(1-\lambda){\frac{\partial pW}{\partial
p}}\rbrace{\nonumber} \\ &&
+\gamma(N+{\frac{1}{2}})\left\{(1+\lambda){\frac{\hbar}{2m\omega_0}}{\frac{{\partial^2}W}{\partial
q^2}}+(1-\lambda){\frac{m\hbar\omega_0}{2}}{\frac{\partial^2
W}{\partial p^2}}\right\},
\label{qle27}
\end{eqnarray} where $\lambda =0$, 1 and -1 correspond to
the master, momentum coupling pre-master and coordinate
coupling pre-master equations, respectively, and where

\begin{equation} 2(N+{\frac{1}{2}})=\coth(\hbar\omega_0
/2kT). \label{qle28}
\end{equation}

As a check, we note that the equilibrium distribution,

\begin{equation}
W_0(q,p)={\frac{1}{(2N+1)\pi}}\exp\lbrace-{\frac{p^2 +m^2
\omega^2_0 q^2}{(2N+1)m\hbar \omega_0}}\rbrace,
\label{qle29}
\end{equation}

satisfies this equation for any value of $\lambda$ if
$f(t)=0$.  It is also instructive to write (\ref{qle27})
explicitly for each value of $\lambda$.

(a) Momentum Coupling Pre-Master $(\lambda=1)$

\begin{eqnarray} {\frac{\partial W}{\partial t}}= &&
-{\frac{p}{m}}{\frac{\partial W}{\partial q}}+m\omega^2_0
q{\frac{\partial W}{\partial p}} -f(t)\frac{\partial
W}{\partial p} \nonumber \\ && +\gamma \frac{\partial
qW}{\partial
q}+\gamma(N+{\frac{1}{2}}){\frac{\hbar}{m\omega_0}}{\frac{\partial^2
W}{\partial q^2}} ~.
\label{qle30}
\end{eqnarray}

(b) Coordinate Coupling Pre-Master $(\lambda =-1)$

\begin{eqnarray} {\frac{\partial W}{\partial t}}= &&
-{\frac{p}{m}}{\frac{\partial W}{\partial
q}}+m\omega^2_0q{\frac{\partial W}{\partial
p}}-f(t)\frac{\partial W}{\partial p} \nonumber \\  && +
\gamma\frac{\partial pW}{\partial p}+\gamma
(N+{\frac{1}{2}})m\hbar\omega_0{\frac{\partial^2
W}{\partial p^2}} ~. \label{qle31}
\end{eqnarray}

(c) Master $(\lambda=0)$

\begin{eqnarray}
\frac{\partial W}{\partial t}= &&
-\frac{p}{m}\frac{\partial W}{\partial
q}+m{\omega^2_0}q\frac{\partial W}{\partial
p}-f(t)\frac{\partial W}{\partial p} {\nonumber}
\\ && +\frac{\gamma}{2}\lbrace\frac{\partial qW}{\partial
q}+\frac{\partial pW}{\partial p}\rbrace{\nonumber}
\\ &&
+\gamma\left(N+\frac{1}{2}\right)\left\{\frac{\hbar}{2m\omega_{0}}~\frac{\partial^{2}
W}{\partial
q^{2}}+\frac{m\hbar\omega_{0}}{2}~\frac{\partial^{2}W}{\partial
p^{2}}\right\} ~.
\label{qle32}
\end{eqnarray}

We note that, in the high-temperature limit, (\ref{qle31})
reduces to the well-known classical Kramers
equation\cite{risken84}, generalized to include
time-dependence in the external force.  Actually,
(\ref{qle30}) is not inconsistent with (\ref{qle31}) since
the values of the momentum $p$ appearing in both equations
are not the same, as discussed in the Introduction (while
emphasizing that $q$ is the same in both equations).  On
the other hand, (\ref{qle32}) does not reduce to the
correct classical result.  Earlier discussions of
different forms of the master and Fokker-Planck equations
may be found in
\cite{dekker77} and \cite{dodonov85}.  In the next
section, we will show how the exact master equation is
obtained from the solution of the initial value quantum
Langevin equation.  As above, it is found to be desirable
to express the result in the form of an equation for the
corresponding Wigner distribution (see (\ref{qle58})
below).

\section{Quantum Langevin Equation: Initial Value Problem}

In this and the following section, we will take $f(t)=0$.
The Langevin equation for the oscillator with given
initial values is given by \cite{ford012}

\begin{equation}
m\ddot{x}+\int^{t}_{0}dt^{\prime}\mu(t-t^{\prime})\dot{x}(t^{\prime})+Kx=-\mu(t)x(0)+F(t),
\label{qle9}
\end{equation} and the general solution is given by

\begin{equation}
x(t)=m\dot{G}(t)x(0)+mG(t)\dot{x}(0)+X(t) \label{qle11}
\end{equation} where we have introduced the fluctuating
position operator,

\begin{equation}
X(t)=\int^{t}_{0}dt^{\prime}G(t-t^{\prime})F(t^{\prime}).
\label{qle39}
\end{equation}

If we assume that at $t=0$ the system is in a state in
which the oscillator is not coupled to the bath and that
the bath is in equilibrium at temperature $T$, we find
that the correlation and commutator are the same as those
for the stationary equation.  Also, the Green function is
the solution of the homogeneous equation,

\begin{equation} m \ddot{G}+\int^{t}_{0}dt^{\prime}\mu
(t-t^{\prime})\dot{G}(t^{\prime})+KG=0, \label{qle40}
\end{equation} with the initial conditions

\begin{equation} G(0)=0,~~~\dot{G}(0)=\frac{1}{m}.
\label{qle41}
\end{equation}

Typically, the memory function $\mu (t)$ falls to zero in
a very short time $\tau$, called the relaxation time of
the bath.  For times long compared with this bath
relaxation time, the extra term on the right hand side of
(\ref{qle9}) vanishes, but only for much longer times,
times long compared with the oscillator decay time, will
this equation become the stationary equation, with the
lower limit on the integration taken to be $-\infty$.  It
was then possible to show \cite{ford012} that one could
write the Langevin equation (\ref{qle9}) in the form of an
equation that is local in time with time-dependent
coefficients:

\begin{equation}
\ddot{x}+2\Gamma(t)\dot{x}+\Omega^{2}(t)x=\frac{1}{m}F(t),
\label{qle42}
\end{equation} where explicit expressions for $\Gamma (t)$
and $\Omega (t)$ were obtained in terms of $G(t)$.
Furthermore, it was shown that these results constitute in
essence a derivation of the HPZ exact master equation
\cite{hpz92} with explicit expressions for the
time-dependent coefficients.

The HPZ equation is perhaps best displayed in the form of
its corresponding Wigner function equation. Using
(\ref{qle26}), we obtained \cite{ford012}

\begin{eqnarray}
\frac{\partial W}{\partial
t}=&&-\frac{1}{m}~p~\frac{\partial W}{\partial
q}+m\Omega^{2}(t)q\frac{\partial W}{\partial p}+2\Gamma
(t)\frac{\partial pW}{\partial p}
\nonumber \\ &&{}+\hbar m\Gamma
(t)h(t)\frac{\partial^{2}W}{\partial p^{2}}+\hbar\Gamma
(t)f(t)\frac{\partial^{2}W}{\partial q\partial p},
\label{qle58}
\end{eqnarray} where $\Omega^{2}(t)$, $2\Gamma (t)$,
$h(t)$, and $f(t)$ are time-dependent parameters for which
we obtained explicit expressions in terms of $X(t)$,
$F(t)$ and $G(t)$.  Since these are rather cumbersome, we
simply refer to equations (2.19) and (3.8) of
\cite{ford012}.  It should also be noted that equations
equivalent to the HPZ equation have been obtained in
\cite{haake85,karrlein97,halliwell96} and in
\cite{caldeira85,unruh89} for the particular case of an
Ohmic bath.

The solution of this exact master equation will be
addressed in the next section.  However, it is prudent to
point out a serious caveat at this stage: a serious
divergence arising from the assumption of an initially
uncoupled state is found to be due to the zero-point
oscillations of the bath and not removed in a cutoff
model.  As a consequence, worthwhile results for the
equation can only be obtained in the high temperature
limit, where zero-point oscillations are
neglected\cite{ford012}.

\section{Wigner Distributions for the Initial Value
Problem}

Our explicit solution (\ref{qle11}) of the Langevin
equation, together with use of Wigner functions enables us
to construct the general solution of the exact master
equation.  Since the latter is an equation for the reduced
density matrix, it was natural to proceed by writing

\begin{equation} W(q,p;t)=\int d{\bf{q}}\int
d{\bf{p}}W_{\rm system}(q,p;{\bf{q}},{\bf{p}};t).
\label{qle43}
\end{equation} Here $W_{\rm system}$ is the Wigner
function for the system of oscillator and bath, with
${\bf{q}}=(q_{1},q_{2}\cdots )$ and
${\bf{p}}=(p_{1},p_{2}\cdots )$ the bath coordinates and
momenta.  Also, since the initial state for the HPZ
equation is a product state, it follows that the
corresponding Wigner function is of the form

\begin{equation} W_{\rm
system}(q,p;{\bf{q}},{\bf{p}};0)=W(q,p;0)\prod_{j}w_{j}(q_{j},p_{j}).
\label{qle44}
\end{equation} Here, on the right $W(q,p;0)$ is the
initial Wigner function for the oscillator and the product
is the Wigner function for the bath, in which
$w_{j}(q_{j},p_{j})$ is the Wigner function for a single
oscillator of mass
$m_{j}$ and frequency $\omega_{j}$,

\begin{eqnarray}
w_{j}(q_{j},p_{j})&=&\frac{1}{\pi\hbar\coth
(\hbar\omega_{j}/2kT)} \nonumber \\
&&{}\times\exp\left\{-\frac{p^{2}_{j}+m^{2}_{j}\omega^{2}_{j}q^{2}_{j}}{m_{j}\hbar\omega_{j}\coth
(\hbar\omega_{j}/2kT)}\right\}. \label{qle45}
\end{eqnarray} After some algebra, we were able to write
the solution in the form of a transition operator acting
on the initial Wigner function,

\begin{equation} W(q,p;t)=\int_{-\infty }^{\infty
}dq^{\prime }\int_{-\infty }^{\infty }dp^{\prime
}P(q,p;q^{\prime },p^{\prime };t)W(q^{\prime },p^{\prime
};0).
\label{qle46}
\end{equation} where $P(q,p;q^{\prime },p^{\prime };t)$,
the transition probability, can be written

\begin{equation}
P(q,p;q^{\prime},p^{\prime};t)=\frac{1}{2\pi\sqrt{\det
{\bf{A}}}}\exp\left\{-\frac{1}{2}{\bf{R}}\cdot \bf{A}^{-1}
\cdot \bf{R} \right\}, \label{qle59}
\end{equation} where we have used a dyadic notation with

\begin{eqnarray} {\bf A}(t) &=&\left(
\begin{array}{cc} m^{2}\left\langle
\dot{X}^{2}\right\rangle & \frac{m}{2}\left\langle
X\dot{X} +\dot{X}X\right\rangle \\
\frac{m}{2}\left\langle X\dot{X}+\dot{X}X\right\rangle &
\left\langle X^{2}\right\rangle
\end{array}
\right) ,  \nonumber \\
{\bf R}(t) &=&\left(
\begin{array}{c} p-\left\langle p(t)\right\rangle \\
q-\left\langle q(t)\right\rangle
\end{array}
\right).  \label{qle60}
\end{eqnarray} Here, in {\bf{R}}, the quantities $\langle
q(t)\rangle$ and $\langle p(t)\rangle$ correspond to the
mean of the initial value solution with initial values
$q^{\prime}$ and $p^{\prime}$.  We refer to
\cite{ford012} for further details.  We also note that
transition probabilities are also discussed in
\cite{dodonov95}.

In fact, the expression (\ref{qle46}) for the transition
probability is formally the same as that for the classical
Kramers equation
\cite{risken84}. The difference is that the Green function
and the mean square of the fluctuating position and
velocity operators here are for a quantum oscillator
interacting with an arbitrary heat bath, while in the
classical solution of the Kramers equation they are for a
classical oscillator interacting with an Ohmic bath.
Also, the classical solution of the Kramers equation
corresponds to taking the stationary limit.

Finally, it should be emphasized that the key results we
obtained in \cite{ford012}, some of which will be
discussed below, depended only on our solution of the
initial value Langevin equation, in the course of which we
showed equivalence with the HPZ equation.

\section{Decoherence}

The problem of decoherence in quantum systems has been of
considerable recent interest
\cite{caldeira851,walls85,zurek91,giulini96,ford014,ford02,ford01,ford012}.
Decoherence refers to the destruction of a quantum
interference pattern and is relevant to the many
experiments that depend on achieving and maintaining
entangled states.  Examples of such efforts are in the
areas of quantum teleportation\cite{zeilinger00}, quantum
information and
computation\cite{bennett95,quantuminfo98}, entangled
states\cite{haroche98}, Schr\"{o}dinger
cats\cite{zeilinger001}, and the quantum-classical
interface\cite{tegmark01}.  For an overview of many of the
interesting experiments involving decoherence, we refer
to Refs.
\cite{haroche98} and
\cite{myatt00}.

Much of the discussion of decoherence has been in terms of
a particle moving in one dimension that is placed in
initial superposition state (a Schr\"{o}dinger "cat"
state) corresponding to two widely separated wave
packets (see (\ref{qle48}) below) separated by a distance
$d$.  Decoherence is said to occur when the long-time
interference pattern is destroyed.  The key questions
asked are, first, under what conditions does decoherence
occur and, second, what is the decoherence time.  In broad
outline, the problem is that of a quantum particle in a
superposition state and interacting with the environment
and two distinct physical scenarios present themselves
depending on whether the particle is initially decoupled
from the environment or, alternatively, in thermal
equilibrium with the environment at the time it is put
into the initial state by a measurement.  We will now
examine each situation in turn, using the results
developed in Sections VI and VII.

\subsection{Initial Decoupling of Particle and
Environment}

This has been the situation which has examined by most
previous investigators and the tools of choice are either
the Feynman-Vernon influence functional techniques
\cite{caldeira851} or master equation techniques
\cite{walls85,zurek91,giulini96,ford012}.  It has now
become clear that, in general, the Lindblad and other
weak-coupling master equations are not adequate because
they cannot properly handle the short times associated
with decoherence phenomena \cite{ford012}.  We have
recently given an exact treatment of the problem based on
our exact solution of the Langevin equation for the
initial value problem and concomitantly, the exact
solution of the HPZ master equation\cite{ford012}.  As
discussed in Sec. VII, the solution is best presented as a
Wigner function for the reduced density matrix.  We recall
that the HPZ equation is a master equation with
time-dependent coefficients for a harmonic oscillator
interacting with a linear passive heat bath of
oscillators.  The equation is exact and general within the
assumption that in the initial state the bath is in
equilibrium and not coupled to the oscillator.  This
assumption of a decoupled initial state is common to all
derivations of a master equation, going back at least to
the work of Wangness and Bloch \cite{wangness53}, who
phrased it as an assumption that at any instant of time
the system is approximately decoupled.  Indeed, such an
assumption is essential for the introduction of the notion
of partial trace, i.e., the trace over states of the
uncoupled bath, key to the existence of any master
equation.  The bath at $t=0$ is in equilibrium at
temperature $T$ whereas, for the oscillator, we have the
freedom to decide its initial state.  More often, the
initial temperature of the bath is taken to be zero or $T$
but it is important to emphasize that, depending on the
choice, different predictions ensue
\cite{ford012}.  Hence, we will examine both situations
in turn.  In both cases, we will need as a starting-point
the probability density at time $t$, given by

\begin{equation}
P(x;t)=\int^{\infty}_{-\infty}dpW(x,p;t), \label{qle47}
\end{equation} where $W(x,p,t)$ is the reduced Wigner
function.  This enabled us to obtain the general result
\cite{ford012}

\begin{eqnarray} P(x;t) &=&
\frac{1}{2\pi\hbar}\int^{\infty}_{-\infty}ds\tilde{W}(Gs,m\dot{G}s;0)
\nonumber \\ && \times
\exp\left\{-\frac{1}{2\hbar^{2}}\langle X^{2}\rangle
s^{2}+i\frac{x}{\hbar}s\right\}. \label{qle61}
\end{eqnarray} where $\tilde{W}$ is the Fourier transform
of the Wigner function.

\subsection*{(i)~~Particle initially at zero temperature}

This is the only case where one can start with a pure
Schr\"{o}dinger superposition ("cat") state, undisturbed
by temperature or environmental effects (albeit difficult
to achieve in practise).  When this state at temperature
zero is suddenly coupled to the heat bath at temperature
$T$, the initial time dependence is then dominated by the
"warming up" of the particle, which occurs on a time scale
of order the decay time
$\gamma^{-1}$.  However, since $\gamma^{-1}$ is generally
very much greater than the decoherence decay time
$\tau_{d}$, decoherence will have occured before the
particle has reached the bath temperature, as a result of
which an important contribution to $\tau_{d}$ is missed.
To see this explicitly, we considered an initial state
corresponding to two separated Gaussian wave packets.  The
corresponding wave function has the form

\begin{equation}
\psi (x,0)=\frac{1}{(8\pi \sigma
^{2})^{1/4}(1+e^{-\frac{d^{2}}{8\sigma ^{2}}%
})^{1/2}}(\exp \{-\frac{(x-\frac{d}{2})^{2}}{4\sigma
^{2}}\}+\exp \{-\frac{
^{2}}\}),  \label{qle48}
\end{equation} where $d$ is the separation and $\sigma $
is the width of each packet.  Next, using (\ref{qle47}),
we obtain an expression for the probability distribution
which consists of three terms, the first two of which
correspond to the individual wave packets whereas the
third is an interference term, which consists of a cosine
term multiplied by a time-dependent factor.  The
attentuation factor $a(t)$ is the ratio of the coefficient
of the cosine term divided by twice the geometric mean of
the first two terms.  We find, for an Ohmic bath, high $T$
and
$t<<\gamma^{-1}$ \cite{ford012}

\begin{equation} a(t)\cong \exp \{-\frac{\zeta
kTd^{2}t^{3}}{12m^{2}\sigma ^{4}+3\hbar ^{2}t^{2}}\},
\label{qle49}
\end{equation} where $\zeta =m\gamma$.  If we suppose that
$\sigma$ is negligibly small, we find $a(t)\cong
\exp \{-t/\tau _{d}\}$ where $\tau _{d}=\frac{3\hbar
^{2}}{\zeta kTd^{2}}$. This, except for a factor of $6$ is
exactly the decoherence time that often appears in the
literature \cite{zurek91,anglin97} for the off-diagonal
terms of the density matrix (as distinct from the spatial
probability discussed here, which is of more interest
because its a measurable quantity). On the other hand, if
the
$\sigma$ term in the denominator dominates, then
decoherence occurs on a time scale $\sim t^{3}$.  But, as
we have seen above, this result corresponds to a particle
in an initial state that is effectively at temperature
zero, which is suddenly coupled to a heat bath at high
temperature. The result is therefore unphysical in the
sense that the initial state does not correspond to that
envisioned when we speak of a system at temperature $T$.

\subsection*{(ii)~~Particle initially at the bath
temperature $T$}

In this case, when the particle and bath are suddenly
coupled, the whole system is immediately in equilibrium at
temperature $T$.  In order to analyze this scenario, we
simply generalize to a state corresponding to a particle
with a random velocity $v$ due to the temperature
environment.  Next, we obtain the corresponding Wigner
function which we average over a thermal environment.
This leads to the result \cite{ford01,ford012}

\begin{equation} a_{{\rm T}}(t)\cong \exp
\{-\frac{\frac{kT}{m}t^{2}}{8(\sigma ^{4}+\sigma
^{2}\frac{kT}{m}t^{2}+\frac{\hbar
^{2}}{4m^{2}}t^{2})}d^{2}\},\qquad t\ll
\gamma^{-1} ,  \label{qle50}
\end{equation} which for very short times is of the form
$a_{{\rm T} }(t)\cong \exp \{-t^{2}/\tau _{d}^{2}\}$,
where the decoherence time is
\begin{equation}
\tau _{d}=\frac{\sqrt{8}\sigma ^{2}}{\bar{v}d},
\label{qle51}
\end{equation} in which $\bar{v}=\sqrt{kT/m}$ is the mean
thermal velocity.  In fact, as we have shown
\cite{ford014,ford02}, the latter result can be obtained
solely within the framework of elementary quantum
mechanics and equilibrium statistical mechanics.  In
particular, since
$\tau_{d}$ is independent of the parameters of the heat
bath, the result given in (\ref{qle50}) exhibits
"decoherence without dissipation"
\cite{ford01,ford014,ford02}.  Thus, very different
results are obtained than for case (i).  Moreover, it is
clear that a similar procedure can be carried out for the
case of a particle initially in a very general state
different from that of zero temperature.

However, as mentioned previously, all of these
calculations are subject to a serious caveat connected
with the assumption of an initially uncoupled state viz.
we found \cite{ford012} that a serious persistent
divergence related to the zero-point oscillations of the
bath and which are not removed in a cutoff model.  As a
consequence, worthwhile results for the equation can only
be obtained in the high temperature limit, where
zero-point oscillations are neglected.  Such problems do
not arise in case (B), which we will now examine.

\subsection{Particle and Environment Entangled at all
times}

In order to describe a state of the system that is
entangled at all times, including the initial time, it is
necessary to abandon master equation methods. Some time
ago, a more general method applicable to such systems was
described by Ford and Lewis \cite{ford86}. In their
method, a system in equilibrium is put into an initial
state (e.g., a wave-packet state) by a measurement and
then at a later time is sampled by a second measurement.
This method of successive measurements has recently been
applied to obtain exact results for the problems of wave
packet spreading and decoherence \cite {ford01}.  For the
decoherence problem, one obtains the result \cite{ford01}

\begin{equation}
a(t)=\exp\left\{-\frac{s(t)}{8\sigma^{2}w^{2}(t)}\right\},
\label{qle52}
\end{equation} for a free particle initially in the state
given by (\ref{qle48}), where $s(t)$ is the mean square
displacement.  For
$f(t)=0$, we have

\begin{eqnarray} s(t) &\equiv &\left\langle [x_{{\rm
s}}(t)-x_{{\rm s}}(0)]^{2}\right\rangle
\nonumber \\ &=&\frac{2\hbar }{\pi }\int_{0}^{\infty
}d\omega {\rm Im}\{\alpha (\omega +i0^{+})\}\coth
\frac{\hbar \omega }{2kT}(1-\cos \omega t),  \label{qle53}
\end{eqnarray} is the mean square displacement for the
stationary process and where the variance is given by

\begin{equation}
w^{2}(t)=\sigma^{2}+s(t)-\frac{[x(t),x(0)]^{2}}{4\sigma^{2}}.
\label{qle54}
\end{equation}

The calculations leading to these results were based on
the quantum probability functions, introduced by Ford and
Lewis\cite{ford86}, and which may be shown to be closely
related to Wigner distribution functions.  In the high
temperature limit, (\ref{qle52}) reduces to the same
result given in (\ref{qle50}).  This is not very
surprising since (\ref{qle50}) corresponds to the particle
and bath both being at the same temperature, albeit
uncoupled, at $t=0$ (case A(ii)).  However, when they are
suddenly coupled, they immediately become entangled, as
distinct from case A(i), where the particle temperature is
initially zero so that it takes a time $\approx
\gamma^{-1}$ for entanglement to occur.

In our view, this scenario is the most physically
realistic one because at all times the particle is coupled
to the bath.  Moreover, we have obtained exact results,
for all $t$, which incorporate both arbitrary temperature
and arbitrary dissipation and, in particular, contains
non-trivial $t$ dependences in both numerator and
denominator, especially in the case where
$kT\approx\hbar\gamma$.  Also, even at $T=0$, writing
$a(t)=\exp\{-b(t)\}$, we find that $b(t)\sim
-t^{2}\log\gamma\tau$ for the single relaxation time model
with
$(t/\tau
)<<1$.\cite{ford012,fordroc}.

While our calculations of
$a(t)$ have concentrated on the time decay of the
interference term in the probability distributions, they
were derived from Wigner distributions or closely related
quantities\cite{ford01,ford012} from which other measures
of decoherence are readily calculated, leading to similar
conclusions.

In order to calculate decoherence in phase space, we start
by considering {\underline{any}} quantum state consisting
of two identical components separated by a distance $d$.
Then, it is not difficult to show that the corresponding
Wigner distribution function, $W^{(2)}(q,p)$ at time
$t=0$, say, is given by

\begin{equation}
W^{(2)}(q,p,0)=N_{0}\left\{W\left(q+\frac{d}{2},p,0\right)+W\left(q-\frac{d}{2},p,0\right)+2\cos
\left(\frac{pd}{\hbar}\right)W(q,p,0)\right\},
\label{qle62}
\end{equation} where $W(q,p,0)$ is the Wigner function for
one of the pairs at $t=0$ and $N_{0}$ is a normalization
factor.  Applying this result to the two-Gaussian
superposition, we concluded that there is no decoherence
in either phase space or momentum space i.e. decoherence
is only manifest in coordinate space \cite{murakami}.

The effect of an external classical force
$f(t)$ on decoherence has also been considered
\cite{ford022}.  In the absence of dissipation and for
negligibly low temperature (which approximates the
experimental conditions discussed in \cite{myatt00}) we
obtained \cite{ford022}

\begin{equation}
a(t)=\exp\left\{-\frac{s_{d}(t)d^{2}}{8\sigma^{2}[\sigma^{2}+s_{d}(t)]}\right\},
\label{qle63}
\end{equation} where

\begin{eqnarray} s_{d}(t) &=& \langle x^{2}_{d}(t)\rangle
\nonumber \\ &=&
\int^{t}_{0}dt^{\prime}\int^{t}_{0}dt^{\prime\prime}G(t-t^{\prime})G(t-t^{\prime\prime})g(
t^{\prime}-t^{\prime\prime}), \label{qle64}
\end{eqnarray} is the mean-square displacement due to the
driving force with

\begin{equation} g(t^{\prime}-t^{\prime\prime})=\langle
f(t^{\prime})f(t^{\prime\prime})\rangle. \label{qle65}
\end{equation} In the case of a random delta-correlated
force so that

\begin{equation} g(t^{\prime}-t^{\prime\prime})=g\delta
(t^{\prime}-t^{\prime\prime}). \label{qle66}
\end{equation} where $g$ is time-independent, we obtained,
for small times characteristic of decoherence phenomena,

\begin{equation}
s_{d}=\frac{gt^{3}}{3m^{2}}(\omega_{0}t<<1). \label{qle68}
\end{equation}

In summary, we have shown that the Wigner distribution
function is an invaluable tool for analyzing dissipative
problems in quantum mechanics including investigations
dealing with decoherence.  Also, with regard to the
latter, we demonstrated that very different results may
ensue depending on the choice of whether the particle is
initially decoupled from the environment or,
alternatively, in thermal equilibrium with the
environment at the time it is put into the intial state by
a measurement.

\section{Acknowledgment}

The thrust of this article is based on joint work carried
out over decades in a very fruitful collaboration with
Professor G. W. Ford.

\end{document}